\documentclass[prb,aps,reprint,twocolumn,showpacs,amsmath]{revtex4-1}

\usepackage{graphicx,color,bm}
\usepackage[english]{babel}
\usepackage{amsmath,amssymb}
\usepackage{stmaryrd,dsfont,txfonts}
\usepackage[sort&compress]{natbib}
\usepackage{subfigure}
\usepackage{mathrsfs}
\usepackage{colordvi}
\usepackage{epsfig}
\usepackage{float}
\usepackage{dcolumn}
\usepackage{ifpdf}

\newcommand{\bfs}[1]{\ensuremath{\boldsymbol{#1}} }

\begin{document}

\title{Spin-orbit interaction in GaAs  wells: From one  to two subbands}
\author{Jiyong Fu}
\thanks{Permanent address: Department of Physics, Qufu Normal University, 273165,
  Qufu, Shandong, China}
\affiliation{Instituto de F\'{\i}sica de S\~ao Carlos, Universidade de S\~ao Paulo, 13560-970 S\~ao Carlos, SP, Brazil}
\author{J. Carlos Egues}
\affiliation{Instituto de F\'{\i}sica de S\~ao Carlos, Universidade de S\~ao Paulo, 13560-970 S\~ao Carlos, SP, Brazil}
\date{\today}
\begin{abstract}
We investigate the Rashba and Dresselhaus spin-orbit (SO) couplings in GaAs quantum wells in the range of well widths $w$ allowing for a transition of the electron occupancy from one to two subbands. By performing a detailed Poisson-Schr\"odinger self-consistent calculation, we determine all the intra- and inter-subband Rashba ($\alpha_1$, $\alpha_2$, $\eta$) and Dresselhaus ($\beta_1$, $\beta_2$, $\Gamma$) coupling strengths.
For relatively narrow wells with only one subband occupied, our results are consistent with the data of Koralek \emph{et al.} [Nature \bfs{458}, 610 (2009)], i.e., the Rashba coupling $\alpha_1$ is essentially independent of $w$ in contrast to the decreasing linear Dresselhaus coefficient  $\beta_1$. When we widen the well so that the second subband can also be populated, we observe that $\alpha_2$ decreases and $\alpha_1$ increases,  both almost linearly with $w$. Interestingly, we find that in the parameter range studied (i.e., very asymmetric wells) $\alpha_2$ can attain zero and change its sign, while $\alpha_1$ is always positive.
In this double-occupancy regime of $w$'s, $\beta_1$ is mostly constant and $\beta_2$ decreases with $w$ (similarly to $\beta_1$ for the single-occupancy regime). On the other hand, the intersubband Rashba coupling strength $\eta$ decreases with $w$ while the intersubband Dresselhaus  $\Gamma$ remains almost constant. We also determine the persistent-spin-helix symmetry points, at which the Rashba and the renormalized (due to cubic corrections) linear Dresselhaus couplings in each subband are equal, as a function of the well width and doping asymmetry. Our results should stimulate experiments probing SO couplings in multi-subband wells.
\end{abstract}

\maketitle

\section{introduction}
The spin-orbit (SO) interaction is a key ingredient in semiconductor
spintronic devices. \cite{Awschalom:2002,zutic:2004}
Most of the proposed schemes for electrical
generation, manipulation and detection of electron spin rely on
it.\cite{Awschalom:2002,zutic:2004} Recently, SO effects \cite{winkler:2003} have attracted renewed interest due to several intriguing states, \emph{e.g.}, the persistent spin helix,\cite{schliemann:2003, bernevig:2006, koralek:2009,salis:2012} and  topological states of matter such as topological insulating phases and Majorana fermions in topological superconductors.\cite{bernevig2:2006,lutchyn:2010,oreg:2010}

In GaAs 2D electron gases there are
two dominant SO contributions: the Rashba\cite{rashba:1984} and the
Dresselhaus\cite{dresselhaus:1955} terms, arising from structural and bulk inversion asymmetries, respectively.
The Rashba coefficient can be tuned with the
doping profile or by using an external bias.\cite{engels:1997,nitta:1997}
The Dresselhaus SO interaction contains both linear and cubic terms, with the linear term mainly depending on the quantum well confinement and the cubic one on the electron density.\cite{ koralek:2009,dettwiler:2014} The SO interaction is usually studied in relatively narrow n-type GaAs/AlGaAs wells,\cite{koralek:2009,dettwiler:2014,nitta:2014}
with electrons occupying only the first subband (``single occupancy''). Recently, wider quantum wells with two populated subbands (``double occupancy'') have also attracted interest both experimentally\cite{hu:2000, bentmann:2012,hernandez:2013} and theoretically. \cite{silva:1997,bernardes:2007,calsaverini:2008,fu:2014} The additional orbital degree of freedom gives rise to interesting physical phenomena, \emph{e.g.}, the intrinsic spin Hall effect,\cite{hernandez:2013} interband-induced band anti-crossings and spin mixing in metallic films,\cite{bentmann:2012} and \emph{crossed} spin helices.\cite{fu:2014}

\begin{figure}[bth!]
\includegraphics[width=8.0cm]{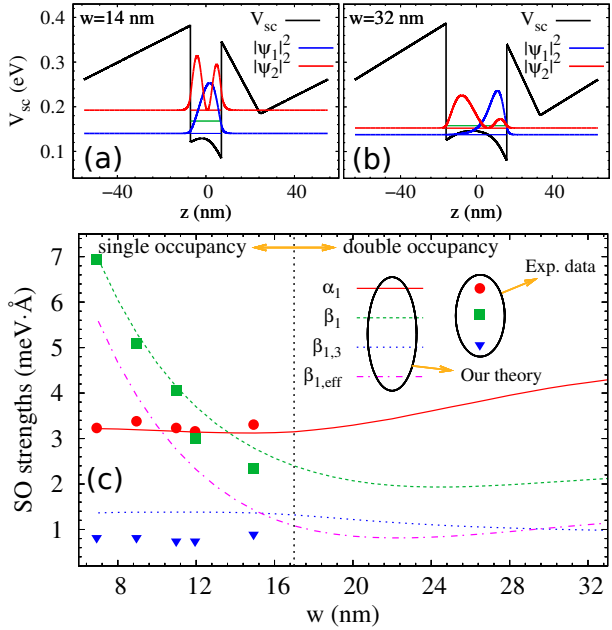}
\caption{(Color online)
Self-consistent potential $V_{sc}$ and wave-function profile $\psi_\nu$ ($\nu=1,2$) in GaAs/Al$_{0.3}$Ga$_{0.7}$As wells,  for a 14-nm well with single occupancy (a) and a 32-nm well  with double occupancy (b).
In (a) $\psi_2$ for the empty second level is also shown for comparison.
The horizontal blue, red and green lines inside the wells indicating the subband energy levels $\mathcal{E}_1$, $\mathcal{E}_2$ and Fermi level $\mathcal{E}_{\rm f}$, respectively. For the14-nm well, $\mathcal{E}_1=140.2$ meV, $\mathcal{E}_2=192.3$ meV, $\mathcal{E}_{\rm f}=168.5$ meV; For the 32-nm well, $\mathcal{E}_1=137.7$ meV, $\mathcal{E}_2=152.7$ meV, $\mathcal{E}_{\rm f}=157.9$ meV.
 (c) Rasha ($\alpha_1$) and Dresselhaus ($\beta_1$, $\beta_{1,3}$, $\beta_{1,\rm eff}=\beta_1-\beta_{1,3}$) coefficients for the first subband as a function of $w$. The markers refers to the experimental data of Ref.\ \onlinecite{koralek:2009} and the curves are from our theory.  The vertical dashed line at $w\sim 17$nm marks a transition from single to double electron occupancy. The total electron density is held fixed at $n_e=8.0\times10^{11}$cm$^{-2}$ as $w$ varies. The temperature is at $75$ K.
}
\label{fig1}
\end{figure}

Here we theoretically investigate the SO couplings in n-type GaAs wells in the range of well widths $w$ allowing for a transition from single to double subband  occupancies. The wells that we consider are similar to the samples experimentally studied by Koralek \emph{et al.}\cite{koralek:2009} (For details see Sec.~\ref{system}). By self-consistently solving the Schr\"odinger and Poisson equations, we determine the confining electron potential and envelope functions, see Figs.\ \ref{fig1}(a) for $w=14$ nm (single occupancy) and \ref{fig1}(b) for $w=32$ nm (double occupancy). We then evaluate the relevant SO strengths, i.e., the intrasubband $\alpha_\nu$ ($\nu=1,2$) and intersubband $\eta$ Rashba couplings and similarly for the Dresselhaus term, the intrasubband $\beta_\nu$ and the intersubband $\Gamma$.
For narrow wells with one subband occupied [left panel in Fig.\ \ref{fig1}(c)], we find that the linear Dresselhaus term $\beta_1$ strongly depends on $w$, while the Rashba $\alpha_1$ is essentially constant, consistent with the data of Koralek \emph{et al.}\cite{koralek:2009} When we widen the well beyond $w\sim 17$ nm  [vertical dashed line in Fig.\ \ref{fig1}(c)] while keeping all other parameters the same, the second subband becomes populated. In this range $\beta_1$ is weakly dependent on $w$, while $\alpha_1$ changes almost linearly [right panel in Fig.\ \ref{fig1}(c)].

For the second subband, the Rashba strength $\alpha_2$ also sensitively depends on $w$; however, it decreases with $w$, in contrast to $\alpha_1$.  Interestingly, our results show that $\alpha_2$ can decrease to zero even for asymmetric wells, further changing its sign [see the arrow in Fig.~\ref{fig2}(b)], while $\alpha_1$ is always greater than zero. This implies that the Rashba couplings of the two subbands can have opposite signs [cf. $\alpha_1$ and $\alpha_2$ in Figs.\ \ref{fig2}(a) and \ref{fig2}(b)] for $w>31$ nm. In addition, the linear Dresselhaus term  $\beta_2$ changes by about a factor of three over the range $w=17-33$ nm, as opposed to $\beta_1$ [cf. Figs.\ \ref{fig1}(c) and \ \ref{fig2}(c)]. As for the the intersubband coefficient $|\eta|$, we find that it decreases with $w$ [Fig.\ \ref{fig2}(d)].
We finally obtain all the persistent-spin-helix symmetry points $\alpha_\nu=\pm\beta_{\nu,\rm eff}$ as a function of $w$ and the doping asymmetry of the wells [Figs.\ \ref{fig3}(a) and \ref{fig3}(b)], where $\beta_{\nu,\rm eff}=\beta_\nu-\beta_{\nu,3}$ is the renormalized ``linear'' Dresselhaus coupling due to cubic corrections ($\beta_{\nu,3}$).  In particular, we are able to identify a unique configuration, $\alpha_1=\beta_{1,\rm eff}$ and  $\alpha_2=-\beta_{2,\rm eff}$, which is crucial for nonballistic spin field effect transistors \cite{schliemann:2003} operating with orthogonal spin quantization axes \cite{nitta:2012} and \emph{crossed} persistent spin helices.\cite{fu:2014}

This paper is organized as follows. In Sec.\ \ref{model hami}, we derive the effective 2D Hamiltonian for electrons in our wells and present the relevant expressions for the intra- and intersubband SO interactions. In Sec.\ \ref{results}, we present our self-consistent results and discussion.  We summarize our main findings in Sec.~\ref{summary}.

\section{Theoretical formulation}
\label{model hami}

Here we outline the derivation of an effective 2D Hamiltonian for electrons in multi-subband quantum wells with SO interactions (Rashba and Dresselhaus). More specifically, we derive an effective 2D Hamiltonian for a two-subband system and determine the relevant SO couplings.

\subsection{From a 3D to an effective 2D Hamiltonian}
The 3D Hamiltonian for electrons in the presence of both the Rashba and Dresselhaus SO interactions in a well grown along the $z||$[001] direction is
\begin{equation}
\mathcal{H^{\rm 3D}}=-\frac{\hbar^2}{2m^*}\frac{\partial ^2}{\partial z^2}+V_{sc}(z)+\frac{\hbar^2(k^2_{x}+k_y^2)}{2m^*} +\mathcal{H_R}+\mathcal{H_D},
\label{eq:h3d}
\end{equation}
where $m^*$ is the electron effective mass and $k_{x,y}$ the electron momentum along the $x||$[001] and $y||$[010] directions.
The potential $V_{sc} (z)$ contains the structural confining potential $V_{\rm w}$ arising from the band offset at the well/barrier interfaces, the external gate potential $V_{\rm g}$, the doping potential $V_{\rm d}$, and the electron Hartree potential $V_{\rm e}$.\cite{bernardes:2007, dettwiler:2014, calsaverini:2008} Note that $V_{sc}$ is calculated self-consistently within the (Poisson-Schr\"odinger) Hartree approximation.
The terms $\mathcal{H_R}$ and $\mathcal{H_D}$ describe the Rashba and Dresselhaus SO interactions, respectively.
The Rashba contribution has the form $\mathcal{H_R}=\eta(z)(k_x\sigma_y-k_y\sigma_x)$ with $\eta(z)=\eta_{\rm w} \partial_zV_{\rm w} + \eta_{\rm H} \partial_z (V_{\rm g}+V_{\rm d}+V_{\rm e})$ determining the Rashba strength and $\sigma_{x,y,z}$ the spin  Pauli matrices. The parameters $\eta_{\rm w}$ and $\eta_{\rm H}$  involve essentially bulk quantities of the well layer. \cite{bernardes:2007, calsaverini:2008, dettwiler:2014} The Dresselhaus term reads $\mathcal{H_D}=\gamma [\sigma_x k_x(k_y^2-k_z^2)+\rm c.c.]$, with $\gamma$ the bulk Dresselhaus parameter and $k_z=-i\partial_z$. Here we take $\gamma$ as that of the well material, since the electrons are mostly confined there. \cite{dettwiler:2014}

Now we follow Ref.~\onlinecite{calsaverini:2008} to obtain an effective 2D Hamiltonian $\mathcal{H}^{\rm 2D}$ from $\mathcal{H}^{\rm 3D}$ in Eq.\ (\ref{eq:h3d}).  We first determine (self-consistently) the spin-degenerate eigenvalues $\mathcal{E}_{{\mathbf k_{\|}}\nu}$=$\mathcal{E}_\nu$ +$\hbar^2k^2_{\|}/2m^*$ and the corresponding eigenspinors $| {\mathbf k_{\|}}\nu\sigma\rangle=| {\mathbf k_{\|}}\nu\rangle \otimes |\sigma\rangle$, $\langle \mathbf{r}| {\mathbf k_{\|}}\nu\rangle=\exp(i{\mathbf k_{\|}\cdot \mathbf r_{\|}})\psi_\nu(z)$, of the well {\it in the absence} of SO interaction. \cite{calsaverini:2008} Here we have defined $\mathcal{E}_\nu$ ($\psi_\nu$), $\nu=1,2,...$,  as the $\nu^{\rm th}$ quantized energy level  (wave function), $\mathbf k_{\|}$ as the in-plane electron wave vector and $\sigma= \uparrow,\downarrow$ as the electron spin component along the $z$ direction. We can then straightforwardly find a (``quasi-2D'') $\mathcal{H}^{\rm 2D}$ by projecting $\mathcal{H^{\rm 3D}}$ [Eq.~(\ref{eq:h3d}), with SO] onto the basis  $\{| {\mathbf k_{\|}}\nu\sigma\rangle \}$. In practice, one considers a truncated set with a finite number $N_0$ of subbands $\nu=1,2,...N_0$. \cite{calsaverini:2008} Next we consider the two-subband case, the one in which we are interested here.

In the coordinate system ${x}_+ || (110)$, ${x}_- || (1\bar{1}0)$ and with the basis ordering
$\Big\{|{\mathbf k_{\|}}1\uparrow \rangle, |{\mathbf k_{\|}}1\downarrow \rangle, |{\mathbf k_{\|}}2\uparrow \rangle, |{\mathbf k_{\|}}2\downarrow \rangle \Big\}$, one has
\begin{equation}
\mathcal{H}^{\rm 2D}=\Big(\frac{\hbar^2k^2}{2m^*}+\mathcal{E}_+\Big)\mathbf{1}\otimes\mathbf{1}-\mathcal{E}_-\tau_z\otimes\mathbf{1}+\mathcal{H_{RD}}, \label{H2D}
\end{equation}
with $\mathcal{E}_\pm=(\mathcal{E}_{2}\pm\mathcal{E}_{1})/2$, $\mathbf{1}$ the 2$\times$2 identity matrix (in both spin and orbital subspaces) and $\tau_{x_+,x_-,z}$ the Pauli (``pseudospin") matrices acting within the orbital subspace.
The term $\mathcal{H_{RD}}$ describes the Rashba and Dresselhaus SO contributions in terms of intra- and inter-subband SO fields  $\rm {\mathbf{B}}_{\rm SO}^\nu$ and $\rm {\mathbf{B}}_{\rm SO}^{12}$, respectively,
\begin{eqnarray}
\mathcal{H_{RD}}=  \frac{1}{2}g\mu_B\sum_{\nu=1,2} \Big[\tau_\nu \otimes \bfs{\sigma} \cdot  \rm \mathbf{B}^{\nu}_{\rm SO} + \tau_{x_+} \otimes \bfs{\sigma} \cdot  \mathbf{B}_{\rm SO}^{12}\Big],
\label{eq:ham12}
\end{eqnarray}
with  $g$ the electron $g$ factor, $\mu_B$ the Bohr magneton, and $\tau_{1,2}=(\mathbf{1} \pm \tau_z)/2$. Explicitly, the intrasubband SO field is
\begin{eqnarray}
 \rm {\mathbf{B}}_{\rm SO}^\nu = \frac{2}{g\mu_B}k&&
 \Big\{\big[(\alpha_\nu-\beta_{\nu,\rm eff})\sin\theta +\beta_{\nu,3} \sin3\theta\big] \mathbf{\hat {x}_+} + \nonumber \\
&& \big[(\alpha_\nu+\beta_{\nu,\rm eff})\cos\theta- \beta_{\nu,3}\cos3\theta\big] \mathbf{\hat {x}_-} \Big\},
\label{eq:sointra}
\end{eqnarray}
and  the intersubband SO field is
\begin{eqnarray}
\rm {\mathbf{B}}_{\rm SO}^{12}=\frac{2}{g\mu_B}k
\big[(\eta-\Gamma)\cos\theta \mathbf{\hat {x}_+} - (\eta+\Gamma)\sin\theta \mathbf{\hat {x}_-} \big],
\label{eq:sointer}
\end{eqnarray}
with $\theta$ the angle between ${\mathbf k}$ and the ${x}_+$ axis.
Below we define the SO coefficients appearing
in Eqs. (\ref{eq:sointra}) and (\ref{eq:sointer}).

\subsection{SO coefficients}
For the two-subband case, the projection procedure leading to Eqs.~(\ref{H2D})--(\ref{eq:sointer}) amounts to calculating the matrix elements of $\mathcal{H^{\rm 3D}}$ [Eq.~(\ref{eq:h3d})] in the truncated basis set $\{| {\mathbf k_{\|}}\nu\sigma\rangle \}$, $\nu=1,2$. In this process we obtain the Rashba and Dresselhaus SO coefficients $\alpha_\nu$, $\beta_\nu$, $\eta$, and $\Gamma$, which are defined in terms of the matrix elements,
  \begin{eqnarray}
 \eta_{\nu \nu^\prime} = \langle \psi_\nu |  \eta_{\rm w} \partial_z V_{\rm w} + \eta_{\rm H} \partial_z (V_{\rm g}+V_{\rm d}+V_{\rm e}) |\psi_{\nu^\prime} \rangle,
\label{eq:eta}
\end{eqnarray}
and
\begin{eqnarray}
 \Gamma_{\nu \nu^\prime}=\gamma\langle \psi_\nu |k^2_z|\psi_{\nu^\prime} \rangle,
\label{eq:gamma}
\end{eqnarray}
with the Rashba coefficients $\alpha_\nu \equiv\eta_{\nu \nu}$, $ \eta \equiv \eta_{12}$ and
the Dresselhaus coefficients $\beta_{\nu} \equiv \Gamma_{\nu \nu}$, $\Gamma \equiv \Gamma_{12}$.
We have also defined a renormalized ``linear" Dresselhaus coupling $\beta_{\nu,\rm eff}=\beta_\nu-\beta_{\nu,3}$ [Eq.~(\ref{eq:sointra})], due to the cubic correction $\beta_{\nu,3}=\gamma k_{\nu}^2/4$, where $k_{\nu}=\sqrt{2\pi n_\nu}$ is the $\nu^{\rm th}$-subband Fermi wave number with $n_\nu$ the $\nu^{\rm th}$-subband occupation.

Note that the Rashba strength $\alpha_\nu$ [Eq.~(\ref{eq:eta}),] can be split into several different constituents, i.e., $\alpha_\nu=\alpha_\nu^{\rm g}+\alpha_\nu^{\rm d}+\alpha_\nu^{\rm e}+\alpha_\nu^{\rm w}$, with $\alpha_\nu^{\rm g}=\eta_{\rm H} \langle \psi_\nu | \partial_z V_{\rm g}|\psi_\nu \rangle$ the gate contribution,  $\alpha_\nu^{\rm d}=\eta_{\rm H} \langle \psi_\nu | \partial_z V_{\rm d}|\psi_\nu \rangle$ the doping contribution,  $\alpha_\nu^{\rm e}=\eta_{\rm H} \langle \psi_\nu | \partial_z V_{\rm e}|\psi_\nu \rangle$ the electron Hartree contribution, and  $\alpha_\nu^{\rm w}=\eta_{\rm w} \langle \psi_\nu | \partial_z V_{\rm w}|\psi_\nu \rangle$ the quantum well structural contribution. For the intersubband Rashba term, one has $\eta=\eta^{\rm g}+\eta^{\rm d}+\eta^{\rm e}+\eta^{\rm w}$ with $\eta^{\rm j}$ ($\rm j=g,d,e,w$), a similar expression to that for $\alpha_\nu^{\rm j}$, except that the matrix elements now are calculated between different subbands. Note that all of the SO coupling contributions above depend on the total self-consistent potential $V_{sc}$ as our wave functions are calculated self-consistently (Sec. \ref{results}).

\section{Results and discussion}
\label{results}
\subsection{System}
\label{system}
We consider [001]-grown GaAs quantum wells of  width $w$ sandwiched between 48 nm Al$_{0.3}$Ga$_{0.7}$As barriers, similar to those experimentally investigated by Koralek {\it et al.} \cite{koralek:2009} Our structures contain two delta-doping (Si) layers on either side of the well, sitting 17 nm away from the well interface, with donor concentrations $n_A$  and $n_B$, respectively.\cite{footnote-sample} We define the doping asymmetry parameter $r=(n_B-n_A)/n_d$, with $n_d=n_A+n_B$. This asymmetry parameter $r$ can be used as a design parameter to
control/tailor the SO strengths in our wells. Note that $r=\pm1$ corresponds to one-sided doped asymmetric wells (either $n_A\neq0$ and $n_B=0$ or vice-versa) while $r=0$ denotes symmetric doping ($n_A=n_B$).
We assume that the areal electron density $n_e$ is equal to $n_d$,\cite{koralek:2009} thus ensuring charge neutrality in the system.
Below (Sec.~\ref{soi}), we first calculate the SO couplings as a function of the well width $w$ for the asymmetric samples of Ref.~\onlinecite{koralek:2009} by taking $r=1$ (i.e., one-side doping: $n_B\neq0$, $n_A=0$),  $T=75$ K, and $n_d=n_B=n_e=8.0\times10^{11}$ cm$^{-2}$. We also consider a wider range of $w$'s (beyond those of Ref.~\onlinecite{koralek:2009}) for which two subbands can be occupied.  We then investigate in detail how the parameters $r$, $T$, and $n_e$ affect our results (Sec. \ref{su2}).

\subsection{Calculated SO couplings}
\label{soi}
We perform a detailed self-consistent calculation by solving the Schr\"odinger and Poisson coupled equations within the Hartree approximation for well widths $w$ first ranging between 7 and 15 nm like the samples in Ref. \onlinecite{koralek:2009}. In this range of $w$'s, all wells have only one subband occupied. We then extend the well widths $w$ to 34 nm, which allows for a transition of the electron occupancy from one subband to two subbands. For wells with even larger widths (e.g., $w \sim 35$ nm), our simulation shows that a  third subband starts to be populated (we do not consider this case here).

Before discussing the calculated SO coefficients, let us first have a look at our self-consistent solutions.
Figures\ \ref{fig1}(a) and \ref{fig1}(b) show the potential profile and wave functions for wells with  $w=14$ nm (single occupancy) and $w= 32$ nm (double occupancy). For comparison, we also show $\psi_2(z)$ for the empty second level of the 14-nm well. Notice that the electronic  Hartree repulsion is more pronounced in the wider well [Fig.~\ref{fig1}(b)]. Hence the electron wave functions $\psi_1(z)$ and  $\psi_2(z)$ tend to localize on opposite sides of wider wells in contrast to narrower wells. The Hartree potential gives rise to a ``central barrier'' which in wider wells  make them effective double wells. These general features of our self-consistent solutions are helpful in understanding the dependence of the SO couplings on $w$, as we discuss next.

In Fig.\ \ref{fig1}(c), we show both the Rashba and the Dresselhaus strengths of the first subband as a function of $w$.
For relatively narrow wells with just one subband occupied,  our calculated SO couplings are in agreement with those obtained via the transient-spin-grating experiments in Ref.~\onlinecite{koralek:2009}.
In this single-occupancy regime, the linear Dresselhaus coupling $\beta_1=\gamma \langle \psi_1|k_z^2 |\psi_1\rangle$  strongly depends on the well confinement $w$ (cf. dashed line and squares).
In contrast, the Rashba coupling $\alpha_1$ remains essentially constant with $w$ (cf. solid line and circles).
As for the $\beta_{1,3}=\gamma \pi n_e/2$ coupling (cf. dotted line and triangles), there seems to be a discrepancy between our calculated values  and the experimental ones. Note, however, that in Ref.\ \onlinecite{koralek:2009} the authors use $\gamma \sim 5$ eV$\cdot$\AA$^3$, obtained by taking  $\langle \psi_1|k_z^2 |\psi_1\rangle=(\pi/w)^2$, which is valid for infinite barriers.  In realistic GaAs/Al$_{0.3}$Ga$_{0.7}$As wells, however, the average of $k_z^2$ can be substantially smaller because of the wave-function penetration into the finite barriers. Together with experimental collaborators,\cite{dettwiler:2014} we have recently performed a thorough investigation on a set of GaAs wells and have found via a realistic fitting procedure (theory and experiment) $\gamma \sim 11.0$ eV$\cdot$\AA$^3$.\cite{dettwiler:2014}  We use this value in our simulations, which is consistent with the value obtained in a recent study by Walser \emph{et al.}\cite{walser:2012} Figure\ \ref{fig1}(c) allows us to determine the persistent-spin-helix symmetry point
$\alpha_1 = \beta_{1,\rm eff}$ at $w=10.3$ nm, also in good agreement with the experimental value $w=11$ nm.   \cite{koralek:2009}

\begin{figure}[bth]
\includegraphics[width=8.3cm]{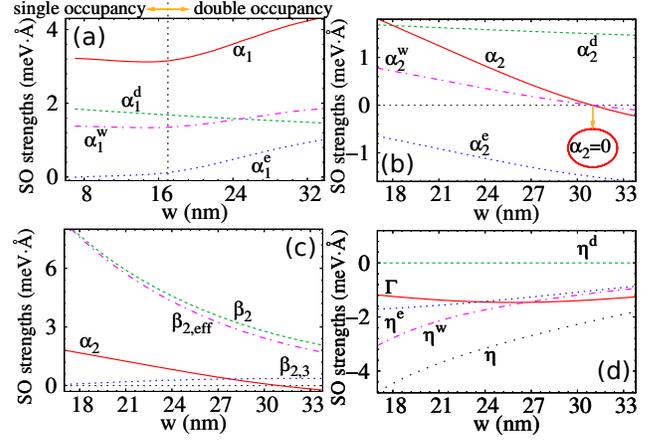}
\caption{(Color online) Distinct contributions to the intrasubband Rashba strength of the first (a) and second (b) subbands, respectively, as a function of $w$. These include the doping contribution $\alpha_\nu^{\rm d}$, the electron Hartree contribution $\alpha_\nu^{\rm e}$, and the structural contribution $\alpha_\nu^{\rm w}$. (c) Intrasubband Rashba ($\alpha_2$) and Dresselhaus ($\beta_2$, $\beta_{2,3}$, $\beta_{2,\rm eff}=\beta_2-\beta_{2,3}$) coefficients for the second subband and (d) intersubband Rashba ($\eta$) and Dresselhaus ($\Gamma$) strengths as a function of $w$. The total electron density is held fixed at $n_e=8.0\times10^{11}$cm$^{-2}$ as $w$ varies. The temperature is at $75$ K. The vertical dashed line in (a) at $w\sim 17$nm marks a transition of the electron occupancy from one subband to two subbands.
}
\label{fig2}
\end{figure}

When we widen the well beyond $w\sim17$ nm, we find that $\beta_{1,3} \propto n_1$ starts to decrease [Fig.\ \ref{fig1}(c)], which indicates a transfer of electrons from the first subband to the second subband. Notice that the total electron density is held fixed at $n_e=n_1+n_2=8.0\times10^{11}$ cm$^{-2}$ as $w$ varies.\cite{footnote-discon} In this double-occupancy regime, the dependence of $\alpha_1$ and $\beta_1$ on $w$ is nearly reversed as compared to the single-occupancy regime. More specifically, here we find that $\beta_1$ remains essentially constant, while $\alpha_1$ changes almost linearly with $w$ [Fig.\ \ref{fig1}(c)].
The new behavior of the SO couplings follows essentially from $\psi_1$ and $\psi_2$ tending to be more localized on opposite sides of wider wells (i.e., $w > 17$ nm) as mentioned before [cf. $\psi_1$ and $\psi_2$ in Fig.~\ref{fig1}(b)]. In this case $\beta_1$ is essentially independent of $w$  because $\psi_1$ is mostly confined to the right half ($z>0$) of the well (in a narrow ``triangular potential'' ) and cannot ``see'' the whole extension $w$ of the well since $\psi_1(z<0)\sim 0$. The linear dependence of $\alpha_1$ on $w$ arises mainly from the electron Hartree  $\alpha_1^{\rm e}$ and the quantum well structural  $\alpha_1^{\rm w}$ contributions, as we discuss next [Fig.\ \ref{fig2}(a)].

Figure \ref{fig2}(a) shows the Rashba couplings $\alpha_1$ and its distinct contributions as a function of the well width $w$. As the well widens, the Hartree contribution $\alpha_1^e$, which is essentially constant and small in the single-occupancy regime, starts to increase almost linearly for $w>17$ nm, as the second subband becomes occupied. The quantum well structural contribution $\alpha_1^w$ presents a similar behavior, but  with $\alpha_1^w > \alpha_1^e$. Both behaviors follow from the already discussed tendency of the envelope subband wave functions $\psi_1$ and $\psi_2$ to localize on opposite sides of the well as $w$ increases, thus making wider wells less symmetric. As both $\alpha_1^e$ and $\alpha_1^w$ are expectations values of the derivatives of the Hartree and structural potential contributions, their corresponding behaviors above follow. The doping contribution $\alpha_1^{\rm d}$ decreases  almost linearly as a function of $w$.  This follows straightforwardly from $V_d(z)\propto z$ within the well, as it arises from a narrow doping region adjacent to the well as explained  in Ref.~\onlinecite{footnote-doping}.

Similarly to $\alpha_1$, the Rashba coupling $\alpha_2$ also changes linearly with $w$. Here the structural, doping and Hartree contributions play similar roles to those for $\alpha_1$. However, $\alpha_2$ decreases with $w$ in contrast to $\alpha_1$, as $\psi_1$ and $\psi_2$ sample  opposite sides of the well for increasing $w$. Surprisingly, the well structural contribution $\alpha_2^w$ is zero for $w\sim 31 $ nm and this implies that $\alpha_2$ is also zero at this $w$, as we show next.

From Ehrenfest's theorem we have
$\langle \partial_z V_{sc}\rangle_\nu= 0=\langle \psi_\nu | \partial_z (V_{\rm w} +V_{\rm g}+V_{\rm d}+V_{\rm e}) |\psi_{\nu} \rangle$ or
$\langle \psi_\nu | \partial_z V_{\rm w}| \psi_\nu\rangle =-\langle\psi_\nu | \partial_z (V_{\rm g}+V_{\rm d}+V_{\rm e}) |\psi_{\nu} \rangle$.
Since  $\alpha_{\nu} = \langle \psi_\nu |  \eta_{\rm w} \partial_z V_{\rm w} + \eta_{\rm H} \partial_z (V_{\rm g}+V_{\rm d}+V_{\rm e}) |\psi_{\nu} \rangle$, we have $\alpha_{\nu} = (\eta_w - \eta_H)\langle \psi_\nu | \partial_z V_{\rm w}| \psi_\nu\rangle$. Now, the structural confining potential of a single quantum well of width $w$ centered at $z=0$ is $V_w \propto [\Theta(w/2 - z) + \Theta(z-w/2)]$, hence $\partial_z V_{\rm w} \propto [- \delta(z+w/2) + \delta(z-w/2)]$ which leads to $\alpha_{\nu} \propto (|\psi_\nu (w/2) |^2-|\psi_\nu (-w/2) |^2)$. Therefore for some particular $w$, $\alpha_\nu$ can in principle be zero provided $|\psi_\nu (w/2)| = |\psi_\nu (-w/2)|$. As shown in Figs.~\ref{fig2}(a) and (b), this can happen for $\alpha_2$ in wider wells -- but not for $\alpha_1$. This, again, follows straightforwardly from the forms of $\psi_1$ and $\psi_2$ in {\it asymmetric} (total potential) wells. As an aside we note that we can alternatively write  $\alpha_{\nu} = - (\eta_w - \eta_H) (\langle\psi_\nu | \partial_z (V_{\rm g}+V_{\rm d}+V_{\rm e}) |\psi_{\nu} \rangle)$. This form shows that when the well structural contribution is zero ($\langle \psi_\nu | \partial_z V_{\rm w}| \psi_\nu\rangle =0$), the corresponding expectation value of $\partial_z(V_g+V_d + V_e)$ also vanishes, as can be seen in Fig.~\ref{fig2}(b) for $w\sim 31$ nm (see arrow). Since we do not consider any gate potential ($V_g=0$), when $\alpha_2^w=0$ then $\alpha_2^d + \alpha_2^e=0$, i.e.,  $\alpha_2^e=-\alpha_2^d$. Note that  $\alpha_1$ and $\alpha_2$ have opposite signs for $w>31$ nm [cf. Figs.\ \ref{fig2}(a) and \ref{fig2}(b)]. The vanishing of $\alpha_2$ discussed above can in principle be used as an handle on how to suppress SO-induced spin relaxation mechanisms,\cite{dyakonov:1971,yafet:1952} for electrons in the second subband.
\begin{figure}[bth]
\includegraphics[width=8.0cm]{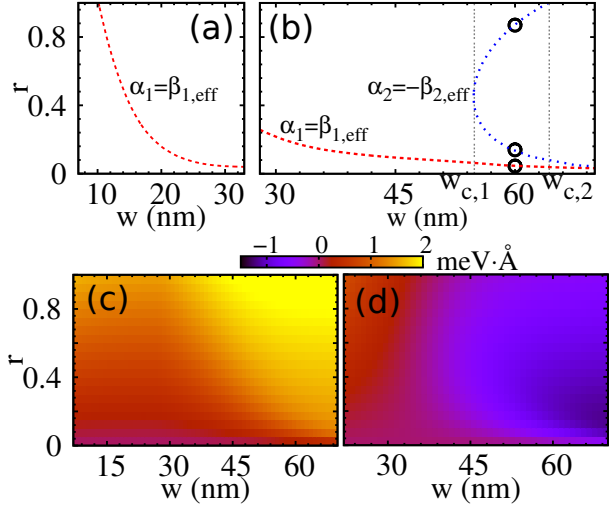}
\caption{(Color online) Symmetry points $\alpha_1=\beta_{1,\rm eff}$ and $\alpha_2=-\beta_{2,\rm eff}$ as a function of the well width $w$ and doping asymmetry parameter $r$, at $n_e=8.0\times10^{11}$ cm$^{-2}$, $T=75$ K (a) and $n_e=4.0\times10^{11}$ cm$^{-2}$, $T=0.3$ K (b). Rashba strength of the first (c) and second (d) subbands as a function of $w$ and $r$ for $n_e=4.0\times10^{11}$ cm$^{-2}$ and $T=0.3$ K. The vertical dashed lines in (b) at $w_{c,1}=55$ nm and $w_{c,2}=64$ nm mark three typical regions of the persistent-spin-helix symmetry points.
}
\label{fig3}
\end{figure}

Figure \ref{fig2}(c) shows the Dresselhaus SO couplings for the second subband ($\alpha_2$ is also shown for comparison). Note that $\beta_2$ and $\beta_{2,{\rm eff}}$ have similar behaviors to those of the corresponding quantities for the first subband in the single-occupancy regime [Fig.~\ref{fig1}(c)]. The coupling $\beta_{2,3} \propto n_2$, however,  increases with $w$ in contrast to $\beta_{1,3}\propto n_1$. This follows from $n_1+ n_2=n_e$ being kept constant in our wells.

Figure \ref{fig2}(d) shows the intersubband Rashba coupling $\eta$ (and its distinct contributions $\eta^{\rm d,e,w}$) and  the Dresselhaus coupling $\Gamma$.
We find that the Rashba strength $|\eta|=|\eta^{\rm d}+\eta^{\rm e}+\eta^{\rm w}|$ decreases with $w$.
This is due to a reducing overlap between $\psi_1$ and $\psi_2$ as $w$ increases [cf. Fig.\ \ref{fig1}(a) and Fig.\ \ref{fig1}(b)].
Since $V_{\rm d}$ is linear (i.e., $\partial_z V_{\rm d}$ is constant) across the well region,\cite{footnote-doping} the doping contribution $\eta^{\rm d}$ is obviously zero due to the orthogonality of $\psi_1$ and $\psi_2$. Therefore the structural contribution $\eta^{\rm w}$ and  the electron Hartree contribution $\eta^{\rm e}$ dominate the behavior of $\eta$ with $w$.
In particular, $\eta^{\rm w} \propto [\psi_1(w/2)\psi_2(w/2)-\psi_1(-w/2)\psi_2(-w/2)]$ fully depends on the overlap of the $\psi_\nu$'s at the well/barrier interfaces being the most sensitive contribution to $\eta$ with $w$.
In contrast, the Dresselhaus coupling $\Gamma \propto  \langle \psi_1|k_z^2|\psi_2\rangle \propto - \langle\psi_1|V_{sc}(z)|\psi_2\rangle $ depends on the overlap of $\psi_\nu$'s across the whole system, thus remaining essentially constant in the parameter range studied. We remark that in wide enough wells with vanishing overlap of the $\psi_\nu$'s, both $\eta$ and $\Gamma$ tend to zero.

\subsection{Persistent-spin-helix symmetry points}
\label{su2}

Let us now consider the interesting possibility of obtaining persistent-spin-helix symmetry points \cite{schliemann:2003,koralek:2009, fu:2014} where the Rashba $\alpha_\nu$ and the renormalized linear Dresselhaus  $\beta_{\nu, \rm eff}$ couplings have equal strengths, i.e., $\alpha_\nu=\pm\beta_{\nu, \rm eff}$ within the respective subbands. \cite{fu:2014}
At these symmetry points, the orientations of the intrasubband SO fields $\mathbf{B}^{1}_{\rm SO}$ and   $\mathbf{B}^{2}_{\rm SO}$ are momentum independent in the absence of the cubic corrections $\beta_{\nu,3}$, see Eq.~(\ref{eq:sointra}). We do not consider the interband contributions $\eta$ and $\Gamma$ any further in this work. These are relevant only near subband crossings, as discussed in Ref.~\onlinecite{fu:2014}. In what follows we exploit the parameter space of our wells by varying both the well width $w$ and the doping asymmetry parameter $r$. We also consider distinct temperatures and electron densities.

 As we determined earlier, the symmetry point $\alpha_1=\beta_{1,{\rm eff}}$ occurs at $w\sim10.3$ nm for $r=1$, $n_e=8.0\times10^{11}$ cm$^{-2}$, and $T=75$ K  [see Sec. \ref{soi} and Fig.~\ref{fig1}(c)]. We have further calculated all symmetry points as a function of the doping asymmetry $r$ and well width $w$ at this density and temperature, thus obtaining a ``line'' on which $\alpha_1=\beta_{1,{\rm eff}}$ [Fig.\ \ref{fig3}(a)], over the $r$ vs. $w$ grayscale map.  However, for the parameters in Fig.\ \ref{fig3}(a) we do not find symmetry points $|\alpha_2|=\beta_{2,\rm eff}$ for the second subband. By lowering the temperature and the electron density to $T=0.3$ K and $n_e=4.0\times10^{11}$ cm$^{-2}$, respectively, we can obtain $|\alpha_1|=\beta_{1,\rm eff}$ and $|\alpha_2|=\beta_{2,\rm eff}$ in a
wide range of well widths, as shown in Fig.\ \ref{fig3}(b). We note that the  third subband  is kept empty even in very wide wells, e.g., $w\sim70$ nm, for these parameters. In Fig.\ \ref{fig3}(b) we can identify three regions separated by the vertical dotted lines at $w=w_{c,1}=55$ nm and $w=w_{c,2}=64$ nm: (i) $w<w_{c,1}=55$ nm for which we find that only the $\alpha_1=\beta_{1,\rm eff}$ symmetry-point line is possible, (ii) $w_{c,1}<w<w_{c,2}$ where three symmetry points $\alpha_1=\beta_{1,\rm eff}$, $|\alpha_2|=\beta_{2,\rm eff}$, and $|\alpha^\prime_2|=\beta_{2,\rm eff}^\prime$ are possible for each $w$ (see the three circles at $w\sim 60$ nm). Here the two sets for the second subband correspond to distinct asymmetry parameters $r$, and (iii) $w>w_{c,2}$ for which again only one set is possible for the second subband. Note that in regions (ii) and (iii)
$\alpha_1=\beta_{1,\rm eff}$, $\alpha_2=-\beta_{2,\rm eff}$.\cite{footnote-bias} This, in principle, allows for the excitation of persistent spins helices with different pitches along orthogonal directions in the two subbands.\cite{fu:2014} Finally, we show grayscale  maps for the Rashba couplings $\alpha_1$ and $\alpha_2$ in Figs.\ \ref{fig3}(c) and \ref{fig3}(d), respectively.  These maps clearly show that, for the parameter range investigated here, $\alpha_1$ is always positive, while $\alpha_2$ is mostly negative.

\subsection{Random Rashba contribution}

In principle, fluctuations of the dopant density can lead to random Rashba couplings.\cite{glazov:2010,morgenstern:2012} To estimate the size of these fluctuations on the calculated SO couplings in our system we follow Refs.~\onlinecite{glazov:2010,glazov:2005,sherman:2003} 
 and write $\sqrt{\langle \alpha^2_R \rangle}={e^2}\xi{\sqrt{\pi n_d}}/{4\pi \epsilon R_d}$ with the subscript $R$ referring to the random contribution. We assume that the fluctuations in the Rashba coupling are the same in both subbands, i.e., $\sqrt{\langle \alpha^2_{1,R} \rangle}$=$\sqrt{\langle \alpha^2_{2,R}\rangle}$, as the electrons see the same random dopant distribution. Here $e>0$ is the electron charge, $\epsilon$ is the dielectric constant, $R_d$ is the distance between the $\delta$-doping layer and well center, and $\xi=\eta_H -\eta_w$.\cite{dettwiler:2014}      
For our wells, $n_d=8.0 \times 10^{11}$ cm$^{-2}$ and $R_d \sim 20$ nm, we find that the variation of the Rashba coupling is around an order of magnitude smaller than the uniform contribution: $\sqrt{\langle \alpha^2_{\nu,R} \rangle}\sim 0.1 \alpha_\nu$. Compared to the third harmonic Dresselhaus coupling, we find $\sqrt{\langle \alpha^2_{\nu,R} \rangle} \sim 0.3 \beta_{3,\nu}$. Notice that here the third harmonic $\beta_{3,\nu}={\gamma\pi n_\nu}/2$ and the uniform Rashba $\alpha_\nu$ are evaluated self-consistently as we discussed above. 

At the persistent-spin-helix symmetry points,  both the random Rashba coupling and the third harmonic Dresselhaus coupling  can destroy the helix. For our system, the relation $\sqrt{\langle \alpha^2_{\nu, R} \rangle} \sim 0.3 \beta_{3,\nu}$ implies that the third harmonic Dressehaus term dominates the decay of the helix. 
More specifically, the spin relaxation rates for the two subbands due to the random Rashba coupling $\Gamma_{R}^\nu$ and the third harmonic Dresselhaus coupling $\Gamma_D^\nu$ are, $\Gamma_{R}^\nu= {8\langle\alpha_{\nu, R}^2\rangle}m^*k_{F,\nu}R_d/{\hbar^3}$ and $\Gamma_{D}^\nu={\gamma^2k_{F,\nu}^6}\tau_P/{4\hbar^2}$,\cite{glazov:2010,meier:1984} respectively.
 Here $k_{F,\nu}$ is the Fermi wave vector for the $\nu$th subband and $\tau_P$ is the momentum relaxation time. For our GaAs wells, 
$\gamma \sim 11.0$ eV $\AA^3$, $k_{F,\nu} \sim 0.2$ nm$^{-1}$, and $\tau_P \sim 1.0$ ps,\cite{dettwiler:2014} we find the ratio $\Gamma_R^\nu/\Gamma_D^\nu\sim 0.03$. 
Although here the random Rashba coupling has a minor effect on our results, we emphasize that this random Rashba contribution could be important in, for instance, symmetric Si/Ge [or GaAs (110)] wells\cite{sherman:2003,zhou:2010} and InSb narrow gap semiconductor wells.\cite{dugaev:2012} In the former case, only the random Rashba coupling contributes to the  D'yakonov-Perel spin relaxation, while in the latter case, a large SO constant can possibly enhance the random Rashba contribution.

\section{Concluding remarks}
\label{summary}

We have performed a detailed and realistic self-consistent calculation for GaAs wells in a wide range of well widths and potential profiles, thus determining all the relevant SO couplings for wells with one and/or two subbands populated. In particular, for narrower wells with only one subband occupied, we have simulated the Rashba and Dresselhaus couplings for the samples experimentally investigated by Koralek {\it et al.} \cite{koralek:2009} and have found very good agreement. We have also determined the symmetry point at which the Rashba and Dresselhaus coefficients are matched. By increasing the well width $w$ beyond the range of the samples in Ref.\ \onlinecite{koralek:2009}, we have investigated the regime in which two subbands are occupied. Interestingly, we find that for wider wells the Rashba coupling $\alpha_2$ can vanish due to Ehnrenfest's theorem even for asymmetric wells, while $\alpha_1$ is always nonzero. This could be important for suppressing the spin-relaxation processes of both the D'yakonov-Perel \cite{dyakonov:1971} and Elliott-Yafet \cite{yafet:1952} types within the second subband. In addition, we have calculated several contributions to the SO couplings due to the structural, Hartree, and doping potentials, thus showing that the magnitudes and signs of the SO couplings follow from the interplay of all these contributions. For very wide wells with varying degrees of potential asymmetry $r$, we find the interesting possibility of tuning the Rashba and Dresselhaus couplings to symmetry points such that $\alpha_1=\beta_{1,\rm eff}$, $|\alpha_2|=\beta_{2,\rm eff}$, and $|\alpha^\prime_2|=\beta_{2,\rm eff}^\prime$ for a given well width $w$ and distinct asymmetry parameter $r$. In this study we kept the electron density fixed while changing the electron occupancy by varying the well width. We point out that the electron occupancy can also be tuned via an external gate at fixed well width, thus widening the scope for new experiments.
Finally, as Rashba and Sherman observed,\cite{sherman:1988} the dependence of the hole SO coupling
on the subband occupation is nontrivial. Additional work is needed to explore features of the SO
coupling for holes.

\begin{acknowledgments}

This work was supported by FAPESP, CNPq, PRP/USP (Q-NANO), and the Natural Science
Foundation of China (Grant No.~11004120). We are grateful to P. H. Penteado for a critical reading of the manuscript.
J.Y.F acknowledges support from Capes (Grant No.~88887.065021/2014­-00) in the later stage of this work and thanks D. R. Candido, W. Wang and X. M. Li for helpful discussions.

\end{acknowledgments}


\begin{thebibliography}{30}
\bibitem{footnote-doping}
The gradient of the doping potential (due to a single delta-doping layer $n_d=n_B$) across the well region is $\partial_z V_d = {e^2n_d} {(l_0-l_b)}/{\epsilon}{(2l_b+w)}$. To first order in ${w}/{2l_b}$ (assuming $w<<2l_b$), this quantity is linear in $w$: $\partial_z V_d \approx {e^2 n_d}{(l_0-l_b)} (1-{w}/{2l_b})/ {2\epsilon}{l_b}$. Here  $e>0$ is the electron charge, $\epsilon$ the dielectric constant (well region), $l_b$ the width of confining barriers, and $l_0$ the distance between the delta-doping layer and the well/barrier interface.
\bibitem{footnote-bias}
In region (ii) (i.e., $w_{c1}<w<w_{c2}$), the persistent-spin-helix symmetry points for the first subband ($\alpha_1=\beta_{1,\rm eff}$) and second subband ($\alpha_2=-\beta_{2,\rm eff}$ or $\alpha^\prime_2=-\beta_{2,\rm eff}^\prime$) correspond to different doping asymmetries $r$ used to control the total potential of the well and hence  the Rashba coupling. In practice, the well symmetry can alternatively be tuned continuously via an external gate bias.
\end{thebibliography}
\end{document}